\documentclass[11pt,a4paper]{article}
\usepackage[top=1.3in, bottom=1.3in, left=1.1in, right=1.1in]{geometry}
\usepackage{authblk} 
\usepackage{amsmath} 
	\numberwithin{equation}{section}
\usepackage{amsfonts} 
\usepackage[hidelinks]{hyperref}
	\hypersetup{colorlinks=true,
				linkcolor=blue,
				urlcolor=blue,
				citecolor=blue}
\usepackage{setspace}
\usepackage[bottom]{footmisc} 
\usepackage{physics}
\usepackage{relsize} 
\usepackage{xparse} 

\newcommand{\inMode}[1]{{\psi_{#1}^{\mathsmaller{\mathrm{in}}}}} 
\newcommand{\inAnOp}[2]{#1_{#2}^{\mathsmaller{\mathrm{in}}}} 
\newcommand{\inCrOp}[2]{#1_{#2}^{\mathsmaller{\mathrm{in}}\dagger}} 
\newcommand{\inVacBra}{\bra{0_{\mathrm{in}}}} 
\newcommand{\inVacKet}{\ket{0_{\mathrm{in}}}} 
\newcommand{\inVacBKet}{\braket{0_{\mathrm{in}}}{0_{\mathrm{in}}}} 

\DeclareDocumentCommand{\F}{ O{} O{} }{F_{\mathsmaller{#1}}^{#2}(\omega)} 
\DeclareDocumentCommand{\R}{ O{} O{} }{\dot{F}_{\mathsmaller{#1}}^{#2}(\omega,\tau)} 

\DeclareMathOperator{\signum}{sgn} 
\DeclareDocumentCommand\sgn{}{\opbraces{\signum}}
\DeclareDocumentCommand\BigO{}{\opbraces{O}} 
\DeclareDocumentCommand\smallO{}{\opbraces{o}} 

\begin{document}

\title{Radiation from a receding mirror: Unruh-DeWitt detector distinguishes a Dirac fermion from a scalar boson}
\author{W. M. H. Wan Mokhtar\thanks{The author's given name is Wan Mohamad Husni. The current address is Universiti Sains Malaysia.}}
\affil{School of Physics, Universiti Sains Malaysia, \\ 11800 USM, Penang, Malaysia \\ wanhusni@usm.my}
\affil{School of Mathematical Sciences, University of Nottingham, \\ Nottingham NG7 2RD, United Kingdom}
\date{\today\thanks{This is a peer-reviewed, un-copyedited version of an article accepted for publication/published in Class. Quantum Grav. \textbf{37} 075011 (2020). IOP Publishing Ltd is not responsible for any errors or omissions in this version of the manuscript or any version derived from it. The Version of Record is available online at doi:10.1088/1361-6382/ab6f0e.}}
\maketitle

\begin{abstract}
It is well known that a receding mirror in Minkowski spacetime can model the formation of a black hole, producing Hawking-like radiation at late times. We ask what an observer would need to do to discern whether the radiation is fermionic or bosonic. Specialising to massless fields in $1+1$ dimensions, we find that an Unruh-DeWitt detector accomplishes this: the late time transition rate of a detector coupled linearly to the scalar density of a spinor field is proportional to the Helmholtz free energy density of a fermionic thermal bath, hence showing a clear sign of Fermi-Dirac statistics, with no counterpart in the response of a detector coupled linearly to a scalar field or its derivative. By contrast, an observer examining just the stress-energy tensor sees no difference between a fermion and a boson, neither at late times nor early.
\end{abstract}

\section{Introduction}

According to general relativity, a sufficiently massive star will undergo gravitational collapse and form a black hole once it runs out of fuel to support itself \cite{Fabbri:2005mw}. By virtue of quantum effects, Hawking showed that an observer far away from the star will see, at late times, thermal radiation with temperature \cite{Hawking:1974sw}
\begin{align} \label{Bekenstein-Hawking temperature}
	T_{\mathrm{BH}} = \frac{\kappa}{2\pi},
\end{align}
where $\kappa$ is the surface gravity of the black hole. This discovery complements Bekenstein's earlier proposals that black holes should have physical entropy and temperature \cite{Bekenstein:1973ur} and that a universe with black holes obey the generalised second law of thermodynamics \cite{Bekenstein:1974ax}.

Two years after Hawking's result, Davies and Fulling showed that a moving mirror in $1+1$ Minkowski spacetime can produce similar thermal radiation provided the mirror follows a trajectory satisfying certain characteristic behaviour at late times \cite{Davies:1977yv}. The similarity provides a strong motivation to study the simple Davies-Fulling model and use it to gain insights on various aspects of black hole physics. For selected references, see \cite{Carlitz:1986nh,Carlitz:1986ng,Chung:1993rf,Hotta:1994ha,Parentani:1995ts,Weinstein:2001kw,Saida:2007ru,Hotta:2013clt,Hotta:2015yla,Chen:2017lum}. A mirror trajectory where the Bogoliubov coefficients describing particle production are in exact correspondence with those of a null shell collapse, for instance, is given in \cite{Good:2016oey}.

In this paper, we analyse the moving mirror model for a massless spinor field. Our aim is to discern what a local observer would observe, particularly when the mirror follows a trajectory that would produce thermal radiation at late times. To this aim, we focus our attention to two local observables. On one hand, we study the stress-energy tensor of the field, which encodes the flow of energy. One the other hand, we study the transition rate of an Unruh-DeWitt detector coupled linearly to the scalar density of the field \cite{Unruh:1976db,DeWitt:1980hx,Takagi:1986kn,Hummer:2015xaa,Louko:2016ptn}. The latter is of special interest following Louko and Toussaint's calculation \cite{Louko:2016ptn}, which shows that the response of such a detector when undergoing a uniform acceleration in full Minkowski spacetime without a mirror contains a Planck factor instead of Fermi-Dirac's. This, hence, raises a question on the detector's ability to distinguish fermions from bosons in other situations, such as the receding mirror spacetime that we will explore. For simplicity, we restrict ourselves to a static detector in this paper.

When the field is in a state corresponding to early time vacuum, we find that the renormalised stress-energy tensor is identical to that of a massless scalar field at all times \cite{Davies:1977yv,Birrell:1982ix,Good:2011}. This implies that an observer will not be able to tell that the radiation is of fermionic in nature by examining only the flow of energy. On the contrary, an observer using an Unruh-DeWitt detector will generically be able to distinguish a Dirac fermion from a scalar boson. As an illustration, we consider a trajectory for which a mirror reflecting a scalar field is known to emit thermal radiation in the far future \cite{Hodgkinson:2013tsa,Juarez-Aubry:2014jba}. For the mirror reflecting a fermion field, we find that the detector's late time transition rate is proportional to the Helmholtz free energy density of fermions in a thermal bath, hence showing a clear sign of Fermi-Dirac statistics.

In Section \ref{Quantum Spinor Field with A Moving Wall}, we start by outlining the setup of our analysis. This includes a discussion of the boundary condition imposed by the mirror. After quantising the field, we renormalise its stress-energy tensor via the point-splitting method. In Section \ref{Unruh-DeWitt Detector}, we first discuss general features of the Unruh-DeWitt detector that interacts with the spinor field. Specialising to a static detector, we then calculate its transition rate and focus on a mirror trajectory where the late time limit satisfies the late time thermality condition of a collapsing star. We derive the late time transition rate and discuss its significance. Finally, we summarise our findings in Section \ref{Discussion}. To assist reading, technical details are deferred to two appendices at the end of the paper.

We emphasise here that we are analysing a fermionic field. One should hence have in mind massless neutrinos instead of photons. Our mirror, in particular, is not a conventional real-world mirror that reflects photons but is virtually invisible to a stream of neutrinos. For this reason, as a reminder, we will use the term \textit{wall} instead of \textit{mirror} to describe the physical boundary for the fermions.

We use an asterisk $^{*}$ to denote complex conjugation. $\BigO(x)$ denotes a quantity such that $\BigO(x)/x$ remains bounded as $x \to 0$ while $\smallO(1)$ indicates a quantity that vanishes in the limit considered. We employ the Einstein summation convention where repeated indices are summed over and work in the natural unit convention where $\hbar = c = k_{\mathrm{B}} = 1$.

\section{Quantum Spinor Field with A Moving Wall} \label{Quantum Spinor Field with A Moving Wall}

Consider a $1+1$ Minkowski spacetime with the metric $\dd{s}^{2} = \dd{t}^{2} - \dd{z}^{2}$ in the standard coordinates $(t,z)$. Null coordinates $u = t - z$ and $v = t + z$ are defined as per usual. Suppose that there exists a wall following a prescribed trajectory defined by $W(u,v) = 0$ where $W(u,v) := v - w(u)$ for some smooth function $w(u)$. We require that the trajectory to be timelike in the sense that the 2-velocity of the wall is timelike, that is having a positive norm in our metric convention, everywhere on the trajectory. Note that, for a Minkowski spacetime that we are considering here, which has no boundaries, \textit{everywhere} does not include the conformal infinity ``endpoints'' when the trajectory is drawn on the corresponding conformal diagram. Note also that, since the trajectory need not be a geodesic, it need not begin and end at the past and future timelike infinities respectively, albeit being a timelike trajectory (see a remark on page 476 of \cite{Carroll:2004st}). However, in order to ensure that the early time state is well-defined, we restrict ourselves to wall trajectories that are asymptotically inertial in the far past and begin at the past timelike infinity. At late times, a similar condition need not hold. Instead, we allow for wall trajectories that are asymptotically null in the $-z$ direction in the far future and may end at the null future infinity. The position $z$ of the wall at any given time $t$ will be denoted $z_{\mathrm{w}}(t)$.

To the right of the wall ($v \geq w(u)$), we consider a two-component spinor field $\psi$ satisfying the Dirac equation
\begin{align}
	i \partial_{t} \psi = - i \alpha \partial_{z} \psi + m \beta \psi,
\end{align}
where $m \geq 0$ is the mass of the field and $\alpha, \beta$ are $2 \times 2$ hermitian anti-commuting Dirac matrices that square to the identity. The Dirac inner product $(\psi_{1},\psi_{2})$ between any two solutions $\psi_{1}$ and $\psi_{2}$ may be evaluated on any spacelike hypersurface satisfying $v \geq w(u)$. Choosing in particular a constant $t$ hypersurface,
\begin{align}
	(\psi_{1},\psi_{2}) = \int_{z_{\mathrm{w}}(t)}^{\infty} [\psi_{1}(t,z)]^{\dagger} \psi_{2}(t,z) \dd{z}.
\end{align}
Following the convention of \cite{Friis:2011yd,Friis:2013eva}, we introduce a spinor basis $\{U_{+},U_{-}\}$ that is orthonormal, in the sense that $U_{+}^{\dagger} U_{-} = U_{-}^{\dagger} U_{+} = 0$ and $U_{+}^{\dagger} U_{+} = U_{-}^{\dagger} U_{-} = 1$, and satisfy
\begin{align} \label{Action of Dirac matrices on U basis}
	\alpha U_{\pm} = \pm U_{\pm}, \qquad \beta U_{\pm} = U_{\mp}.
\end{align}
In terms of this basis, the field $\psi$ can be expanded as $\psi = \psi_{+} U_{+} + \psi_{-} U_{-}$. For the case of a massless field $m = 0$ which we now specialise to, $\psi_{+}$ and $\psi_{-}$ are the right-mover and left-mover respectively. Hence, we have $\psi_{+} = \psi_{+}(u)$ and $\psi_{-} = \psi_{-}(v)$.

On the surface of the moving wall, we require that $\psi$ satisfy the MIT bag boundary condition $i n_{\mu} \gamma^{\mu} \psi(u,w(u)) = \psi(u,w(u))$ where $\gamma^{\mu} = \{\beta,\beta\alpha\}$ are the Dirac gamma matrices and $n_{\mu}$ is the inward-directed unit normal to the wall \cite{Chodos:1974je}. Due to the way we parametrise $W(u,v)$, that is as $v - w(u)$ instead of $w(u) - v$, we have
\begin{align} \label{Unit Normal}
	n_{\mu}
		= \frac{ \partial_{\mu}{W} }{ \sqrt{ \abs{ \eta^{\alpha\beta} \partial_{\alpha}{W} \partial_{\beta}{W} } } }
		= \frac{1}{2 \sqrt{w'(u)}} ( 1 - w'(u) , 1 + w'(u) ),
\end{align}
where the prime in $w'(u)$ indicates a derivative of $w(u)$ with respect to its argument. The direction of $n_{\mu}$ as given in \eqref{Unit Normal} can be verified by considering, for instance, a wall which is static at the origin in the far past. During the early time of such a scenario, we have $w(u) = u$ and $n_{\mu} = (0,1)$, that is rightward, hence inward, directed. Substituting \eqref{Unit Normal} into the boundary condition and using the fact that a massless field propagates on a null geodesic, we have
\begin{align}
	\psi_{+}(u) = - i \sqrt{ w'(u) } \psi_{-}(w(u))
\end{align}	
everywhere to the right of the wall. Note that the reflected field is not only sensitive to the existence of, but also to the instantaneous motion of the wall through $w'(u)$. In particular, $w'(u)$ directly influence the reflected field's amplitude. This phenomenon, which is not observed in the scalar field case, plays an important role in ensuring, for instance, the convergence of the renormalised stress-energy tensor below.

\subsection{Quantisation}

For a moving wall with inertial motion in the far past, we only need to consider an in-mode ansatz $\inMode{k}$ whose left moving part is proportional to $e^{ - i k v } U_{-}$, where $k \in \mathbb{R} \setminus \{0\}$. The boundary condition on the wall's surface then implies that
\begin{align} \label{In-Mode Solution}
	\inMode{k}(u,v) = - i \sqrt{\frac{w'(u)}{2\pi}} e^{ - i k w(u) } U_{+} + \frac{1}{\sqrt{2\pi}} e^{ - i k v } U_{-},
\end{align}
where a normalisation choice has been made so that the mode function above is normalised in the Dirac inner product such that $(\inMode{k},\inMode{k'}) = \delta(k-k')$ for any $k, k' \in \mathbb{R} \setminus \{0\}$.

A general solution to the Dirac equation can then be expanded as
\begin{align} \label{General Solution}
	\psi(u,v) = \int_{0}^{\infty} \dd{k} \left( \inAnOp{a}{k} \inMode{k}(u,v) + \inCrOp{b}{k} \inMode{-k}(u,v) \right),
\end{align}
where $\inAnOp{a}{k}, \inAnOp{b}{k}$ and $\inCrOp{a}{k}, \inCrOp{b}{k}$ are the annihilation and creation operators respectively with non-vanishing anticommutators
\begin{align} \label{Anti-commutators}
	\{\inAnOp{a}{k},\inCrOp{a}{k'}\}
		= \{\inAnOp{b}{k},\inCrOp{b}{k'}\}
		= \delta(k-k')	\qquad \text{for } k, k' > 0.
\end{align}
These operators define a normalised in-vacuum state $\inVacKet$ satisfying $\inVacBKet = 1$ and
\begin{align} \label{Definition of in-vacuum}
	\inAnOp{a}{k} \inVacKet = \inAnOp{b}{k} \inVacKet = 0 \qquad \text{for } k > 0.
\end{align}
In this paper, we will assume that the field is in this in-vacuum state.

\subsection{Field Propagators}

In the convention of \cite{Takagi:1986kn,Louko:2016ptn}, the positive and negative frequency propagators are defined as
\begin{align}
	S_{ab}^{+}(u,v;u',v')
		&:= \inVacBra \psi_{a}(u,v) \overline{\psi}_{b}(u',v') \inVacKet,
			\label{Positive Frequency Propagator} \\
	S_{ab}^{-}(u,v;u',v')
		&:= \inVacBra \overline{\psi}_{b}(u',v') \psi_{a}(u,v) \inVacKet,
			\label{Negative Frequency Propagator}
\end{align}
respectively where $\overline{\psi} = \psi^{\dagger}\beta$ is the Dirac conjugate of $\psi$. The subscript $a$ in $\psi_{a}$, for instance, denotes the $a$-th component of the two-component spinor $\psi$. Substituting the general solution \eqref{General Solution} and using \eqref{In-Mode Solution}, we obtain
\begin{align}
	&S_{ab}^{+}(u,v;u',v') \nonumber \\
		&\ = \frac{1}{2\pi}
			\int_{0}^{\infty} \dd{k} \bigg( \sqrt{w'(u)w'(u')} e^{- i k (w(u) - w(u'))} U_{+,a} U_{+,c}^{\dagger}
			- i \sqrt{w'(u)} e^{- i k (w(u) - v')} U_{+,a} U_{-,c}^{\dagger}
				\nonumber \\
		&\qquad \qquad \qquad \qquad
			+ i \sqrt{w'(u')} e^{- i k (v - w(u'))} U_{-,a} U_{+,c}^{\dagger}
			+ e^{- i k (v - v')} U_{-,a} U_{-,c}^{\dagger} \bigg) \beta_{cb},
				\label{Splus} \\
	&S_{ab}^{-}(u,v;u',v') \nonumber \\
		&\ = \frac{1}{2\pi}
			\int_{0}^{\infty} \dd{k} \bigg( \sqrt{w'(u)w'(u')} e^{i k (w(u) - w(u'))} U_{+,a} U_{+,c}^{\dagger}
			- i \sqrt{w'(u)} e^{i k (w(u) - v')} U_{+,a} U_{-,c}^{\dagger}
				\nonumber \\
		&\qquad \qquad \qquad \qquad
			+ i \sqrt{w'(u')} e^{i k (v - w(u'))} U_{-,a} U_{+,c}^{\dagger}
			+ e^{i k (v - v')} U_{-,a} U_{-,c}^{\dagger} \bigg) \beta_{cb}.
				\label{Sminus}
\end{align}
We interpret $S_{ab}^{+}$ and $S_{ab}^{-}$ as distributions in the sense of $\epsilon \to 0_{+}$ via the prescription \cite{Birrell:1982ix,Mukhanov:2007zz}
\begin{align} \label{Distributional Interpretation}
	\int_{0}^{\infty} \dd{p} e^{\pm i p z}
		\to \lim_{\epsilon \to 0_{+}} \int_{0}^{\infty} \dd{p} e^{\pm i p z - p \epsilon}
		= \lim_{\epsilon \to 0_{+}} \frac{\pm i}{z \pm i \epsilon}.
\end{align}
This gives us
\begin{align}
	S_{ab}^{+}(u,v;u',v')
		&= - \frac{i}{2\pi} \frac{\sqrt{w'(u)w'(u')}}{w(u) - w(u') - i \epsilon} U_{+,a} U_{+,c}^{\dagger} \beta_{cb}
			- \frac{1}{2\pi} \frac{\sqrt{w'(u)}}{w(u) - v' - i \epsilon} U_{+,a} U_{-,c}^{\dagger} \beta_{cb}
				\nonumber \\
		&\qquad
			+ \frac{1}{2\pi} \frac{\sqrt{w'(u')}}{v - w(u') - i \epsilon} U_{-,a} U_{+,c}^{\dagger} \beta_{cb}
			- \frac{i}{2\pi} \frac{1}{v - v' - i \epsilon} U_{-,a} U_{-,c}^{\dagger} \beta_{cb},
				\label{Splus Distribution} \\
	S_{ab}^{-}(u,v;u',v')
		&= \frac{i}{2\pi} \frac{\sqrt{w'(u)w'(u')}}{w(u) - w(u') + i \epsilon} U_{+,a} U_{+,c}^{\dagger} \beta_{cb}
			+ \frac{1}{2\pi} \frac{\sqrt{w'(u)}}{w(u) - v' + i \epsilon} U_{+,a} U_{-,c}^{\dagger} \beta_{cb}
				\nonumber \\
		&\qquad
			- \frac{1}{2\pi} \frac{\sqrt{w'(u')}}{v - w(u') + i \epsilon} U_{-,a} U_{+,c}^{\dagger} \beta_{cb}
			+ \frac{i}{2\pi} \frac{1}{v - v' + i \epsilon} U_{-,a} U_{-,c}^{\dagger} \beta_{cb},
				\label{Sminus Distribution}
\end{align}
where the limit $\epsilon \to 0_{+}$ is implied. We note here that, in the absence of a wall, the propagators consist only of the first and fourth terms of each expression with $w(x) = x$.

\subsection{Stress-Energy Tensor}

Recall that the stress-energy tensor for a Dirac field is given by \cite{Birrell:1982ix}
\begin{align} \label{Tmunu}
	T_{\mu\nu}
		= \frac{i}{2} \left[ \overline{\psi} \gamma_{(\mu} \partial_{\nu)} \psi
			- (\partial_{(\mu} \overline{\psi}) \gamma_{\nu)} \psi \right],
\end{align}
where $\gamma^{\mu} = \{\beta,\beta\alpha\}$ are the Dirac gamma matrices, $\gamma_{\mu} = \eta_{\mu\nu} \gamma^{\nu}$ are the covariant Dirac gamma matrices and $A_{(\mu\nu)} = (1/2) (A_{\mu\nu} + A_{\nu\mu})$ is the symmetric part of $A_{\mu\nu}$. Upon point-splitting, we have
\begin{align}
	T_{\mu\nu}(t,z)
		= \lim_{t',z' \to t,z} \frac{i}{2} \left[ \overline{\psi}(t,z) \gamma_{(\mu} \partial'_{\nu)} \psi(t',z')
			- (\partial_{(\mu} \overline{\psi}(t,z)) \gamma_{\nu)} \psi(t',z') \right],
\end{align}
where the prime in $\partial'_{\nu}$ indicates a derivative with respect to $t'$ and $z'$, as opposed to $t$ and $z$. Using \eqref{General Solution}, \eqref{Anti-commutators}, \eqref{Definition of in-vacuum} and \eqref{Negative Frequency Propagator}, we can formally expressed the in-vacuum expectation value $\expval{T_{\mu\nu}} \equiv \inVacBra T_{\mu\nu} \inVacKet$ as
\begin{align}
	\expval{T_{tt}}
		&= \lim_{u',v' \to u,v} \frac{i}{2}
			\big( \partial_{u} F(u,v;u',v') + \partial_{v} F(u,v;u',v') \big),
				\label{Ttt} \\
	\expval{T_{zz}}
		&= \lim_{u',v' \to u,v} \frac{i}{2}
			\big( \partial_{u} G(u,v;u',v') - \partial_{v} G(u,v;u',v') \big),
				\label{Tzz} \\
	\expval{T_{tz}} = \expval{T_{zt}}
		&= \lim_{u',v' \to u,v} \frac{i}{4}
			\big( - \partial_{u} F(u,v;u',v') + \partial_{v} F(u,v;u',v') \nonumber \\
		&\qquad \qquad \qquad \quad
			- \partial_{u} G(u,v;u',v') - \partial_{v} G(u,v;u',v') \big),
				\label{Ttz}
\end{align}
where
\begin{align}
	F(u,v;u',v')
		&:= \Tr{S^{-}(u,v;u',v')\beta} - \Tr{S^{-}(u',v';u,v)\beta},
			\label{F of stress-energy tensor} \\
	G(u,v;u',v')
		&:= \Tr{S^{-}(u,v;u',v')\beta\alpha} - \Tr{S^{-}(u',v';u,v)\beta\alpha},
			\label{G of stress-energy tensor}
\end{align}
and the trace is over spinor indices.

Note that, due to the fact that $F(u,v;u',v')$ and $G(u,v;u',v')$ above being functions of the trace of $S^{-}\beta$ and $S^{-}\beta\alpha$ respectively, the second and third terms of \eqref{Sminus}, or equivalently \eqref{Sminus Distribution}, do not contribute to the stress-energy in-vacuum expectation value $\expval{T_{\mu\nu}}$. This can be seen as follows. Since similar arguments apply to both $S^{-}\beta$ and $S^{-}\beta\alpha$, let us take the latter as an example. When we take the spinorial trace, we obtain that the contributions from the second and third terms of \eqref{Sminus Distribution} are proportional to
\begin{align}
\Tr{U_{+} U_{-}^{\dagger} \beta \beta \alpha}
	= \Tr{\alpha U_{+} U_{-}^{\dagger}}
	= \Tr{+ U_{+} U_{-}^{\dagger}}
	= + \sum_{a=1}^{2} U_{+,a} U_{-,a}^{\dagger} = 0, \\
\Tr{U_{-} U_{+}^{\dagger} \beta \beta \alpha}
	= \Tr{\alpha U_{-} U_{+}^{\dagger}}
	= \Tr{- U_{-} U_{+}^{\dagger}}
	= - \sum_{a=1}^{2} U_{-,a} U_{+,a}^{\dagger} = 0,
\end{align}
respectively. In first equality, we have used the cyclic permutation property of a trace and the fact that $\beta$ squares to unity. Then, we applied the properties \eqref{Action of Dirac matrices on U basis} in the second equality and used of the orthogonality condition $U_{\pm}^{\dagger} U_{\mp} = \sum_{a=1}^{2} U_{\pm,a}^{\dagger} U_{\mp,a} = 0$ in the last equality.

Subtracting from \eqref{Ttt} -- \eqref{Ttz} the full Minkowski spacetime vacuum contribution before taking the coincidence limit $u',v' \to u,v$, we obtain from \eqref{Sminus Distribution} that
\begin{align} \label{Renormalised Stress-Energy Tensor Result}
	\expval{T_{tt}}_{\mathrm{ren}}
		= \expval{T_{zz}}_{\mathrm{ren}}
		= - \expval{T_{tz}}_{\mathrm{ren}}
		= - \expval{T_{zt}}_{\mathrm{ren}}
		= - \frac{1}{24\pi} (Sw)(u),
\end{align}
where the subscript ``ren'' indicates renormalisation and
\begin{align} \label{Schwarzian derivative}
	(Sw)(u)
		= \frac{w'''(u)}{w'(u)} - \frac{3}{2} \left( \frac{w''(u)}{w'(u)} \right)^{2}
\end{align}
is the Schwarzian derivative of $w(u)$ with respect to $u$. Note that, to arrive at \eqref{Renormalised Stress-Energy Tensor Result} -- \eqref{Schwarzian derivative}, one has to also take the limit $\epsilon \to 0$ before taking the coincidence limit. Otherwise, $\expval{T_{\mu\nu}}_{\mathrm{ren}}$ will be proportional to $(1-w'(u)^{2})\epsilon^{-2}$, which is divergent as $\epsilon \to 0$.

Upon comparison, we find that every component of the renormalised expectation values above is identical to that of a massless scalar field in a moving mirror spacetime at all times \cite{Davies:1977yv,Birrell:1982ix,Good:2011}. This implies that an observer examining only the stress-energy tensor will not be able to discern that the radiation is made up of fermions.

On this note, we would like to remark that, even though the stress-energy tensor does not have the spin-statistics information of the radiation, the canonical spin-statistics connection could be derived from the dynamics using the Bogoliubov coefficients as has been shown in \cite{Good:2012cp} for a scalar field. We have not, however, verified this expectation.

\section{Unruh-DeWitt Detector} \label{Unruh-DeWitt Detector}

We consider now a point-like Unruh-DeWitt detector with two energy levels $\ket{0_{D}}$ and $\ket{\omega}$, associated to energies $0$ and $\omega$ respectively, following a smooth timelike trajectory $(u(\tau),v(\tau))$ where $\tau$ is the detector's proper time \cite{Unruh:1976db,DeWitt:1980hx,Birrell:1982ix}. To ensure that the detector is on the same side as the field $\psi$ and never collides with the wall even in the asymptotic past and future, we require that the condition $v(\tau) - w(u(\tau)) \geq h_{\mathrm{min}}$ holds for all $\tau \in \mathbb{R}$, where $h_{\mathrm{min}}$ is a real positive constant. We will refer to this constraint as the no-collision condition. For a given $w(u)$, the requirement puts a restriction on the detector's trajectory. Conversely, a given detector trajectory limits our choice of $w(u)$.

\subsection{The Response Function}

We couple our detector to the spinor field via the interaction Hamiltonian \cite{Takagi:1986kn,Hummer:2015xaa,Louko:2016ptn}
\begin{align} \label{Interaction Hamiltonian}
	H_{\mathrm{int}} = c \mu(\tau) \chi(\tau) \overline{\psi}(u(\tau),v(\tau)) \psi(u(\tau),v(\tau)),
\end{align}
where $c$ is the coupling constant, $\mu(\tau)$ is the detector's monopole moment and $\chi(\tau)$ is a smooth switching function which specifies how the interaction is switched on and off \cite{Satz:2006kb}. We assume that $\chi$ takes non-negative real values and has a compact support.

Suppose that the detector is prepared in the state $\ket{0_{D}}$ and the field is in the state $\inVacKet$ before the interaction is turned on. Working within first order perturbation theory, the probability $P(\omega)$ for the detector to make a transition to the state $\ket{\omega}$ and the field to any state after the interaction ceased factorises as \cite{Birrell:1982ix}
\begin{align}
	P(\omega)
		= c^{2} \abs{\bra{0_{D}} \mu(0) \ket{\omega}}^{2} F(\omega),
\end{align}
where the pre-factor $\abs{\bra{0_{D}} \mu(0) \ket{\omega}}^{2}$ depends only on the internal structure of the detector. All dependence on the field's initial state, the detector's trajectory and the switching function are encoded in the response function $F(\omega)$ which reads
\begin{align} \label{Response Function}
	\F = \int_{-\infty}^{\infty} \dd{\tau} \int_{-\infty}^{\infty} \dd{\tau'} e^{-i\omega(\tau-\tau')}
			\chi(\tau) \chi(\tau') W^{(2,\bar{2})}(\tau,\tau'),
\end{align}
where $W^{(2,\bar{2})}(\tau,\tau') = W^{(2,\bar{2})}(u(\tau),v(\tau);u(\tau'),v(\tau'))$ is the pull-back of the two-point correlation function
\begin{align}
	W^{(2,\bar{2})}(u,v;u',v')
		&:= \inVacBra \overline{\psi}(u,v) \psi(u,v) \overline{\psi}(u',v') \psi(u',v') \inVacKet
\end{align}
onto the detector's worldline. For this reason, in an abuse of terminology, we will now drop the pre-factor and follow the common convention of occasionally referring to $F(\omega)$ as the transition probability.

For our purposes, it is more convenient to express the response function as \cite{Satz:2006kb,Schlicht:2003iy}
\begin{align} \label{Causal Response Function}
	\F = 2 \int_{-\infty}^{\infty} \dd{r} \chi(r) \int_{0}^{\infty} \dd{s} \chi(r-s)
			\Re{e^{- i \omega s} W^{(2,\bar{2})}(r,r-s)}
\end{align}
using the properties $W^{(2,\bar{2})}(\tau,\tau') = [W^{(2,\bar{2})}(\tau',\tau)]^{*}$. The main advantage of the form above is that it expresses the following causality structure of the detector's response. The transition probability of the detector after all interaction ceased is given by the sum of contributions from every time $r$ where $\chi$ is non-vanishing, that is, when the detector interacts with $\psi$. The contribution at each time $r$ in turn is given by the sum of correlations between $\overline{\psi}\psi$ at the time $r$ and $\overline{\psi}\psi$ in the past, back until when the detector-field interaction is switched on.

Using \eqref{General Solution} -- \eqref{Negative Frequency Propagator}, we may write the two-point function in terms of the positive and negative frequency propagators as \cite{Louko:2016ptn}
\begin{align} \label{Two-point function in terms of propagators}
	W^{(2,\bar{2})}(u,v;u',v')
		&= \Tr{S^{+}(u,v;u',v')S^{-}(u',v';u,v)} \nonumber \\
		&\qquad + \Tr{S^{-}(u,v;u,v)}\Tr{S^{-}(u',v';u',v')}.
\end{align}
Similar to the propagators in which it is expressed in above, $W^{(2,\bar{2})}$ should be interpreted as a distribution. In Appendix \ref{Regularisation of Two-Point Function}, we argue that $W^{(2,\bar{2})}$ may be understood, as a distribution, as the $\epsilon \to 0_{+}$ limit of
\begin{align} \label{Regularised Two-Point Function}
	W_{\epsilon}^{(2,\bar{2})}
		&= - \frac{1}{2\pi^{2}} \bigg[ \frac{ \sqrt{w'(u)w'(u')} }{ ( w(u) - w(u') - i \epsilon ) ( v - v' - i \epsilon ) }
			- \frac{ \sqrt{w'(u)w'(u')} }{ ( w(u) - v' - i \epsilon ) ( v - w(u') - i \epsilon ) }
				\nonumber \\
		&\qquad \qquad \qquad
			- 2 \frac{ \sqrt{w'(u)} (w(u) - v) }{ ((w(u) - v)^{2} + \epsilon^{2}) }
							\frac{ \sqrt{w'(u')} (w(u') - v') }{ ((w(u') - v')^{2} + \epsilon^{2}) } \bigg].
\end{align}
The response function $\F$ in turn, be it in the form of \eqref{Response Function} or \eqref{Causal Response Function}, should be evaluated as follows: make the replacement $W^{(2,\bar{2})} \to W_{\epsilon}^{(2,\bar{2})}$, then perform the double integral and finally take the limit $\epsilon \to 0_{+}$. This is to say that
\begin{align} \label{Detector Response}
	\F = \F[][(0)] + \F[][(1)] + \F[][(2)],
\end{align}
where
\begin{align}
	\F[][(0)]
		&= \lim_{\epsilon \to 0_{+}} - \frac{1}{\pi^{2}} \int_{-\infty}^{\infty} \dd{r} \chi(r)
			\int_{0}^{\infty} \dd{s} \chi(r-s)
				\nonumber \\
		&\hspace{4em} \times
			\Re{ \frac{ e^{- i \omega s} \sqrt{w'(u(r))w'(u(r-s))} }
				{ (w(u(r)) - w(u(r-s)) - i \epsilon)
					( v(r) - v(r-s) - i \epsilon ) } }, \\
	\F[][(1)]
		&= \lim_{\epsilon \to 0_{+}} \frac{1}{\pi^{2}} \int_{-\infty}^{\infty} \dd{r} \chi(r)
			\int_{0}^{\infty} \dd{s} \chi(r-s)
				\nonumber \\
		&\hspace{4em} \times
			\Re{ \frac{ e^{- i \omega s} \sqrt{w'(u(r))w'(u(r-s))} }
				{ (w(u(r)) - v(r-s) - i \epsilon) ( v(r) - w(u(r-s)) - i \epsilon ) } },
				\label{FMW1} \\
	\F[][(2)]
		&= \frac{2}{\pi^{2}} \int_{-\infty}^{\infty} \dd{r} \chi(r) \frac{ \sqrt{w'(u(r))} }{ w(u(r)) - v(r) }
			\int_{0}^{\infty} \dd{s} \chi(r-s) \frac{ \cos(\omega s) \sqrt{w'(u(r-s))} }{ w(u(r-s)) - v(r-s) }.
				\label{FMW2}
\end{align}
In arriving at \eqref{FMW2}, we have used the no-collision condition and dominated convergence theorem to justify setting $\epsilon = 0$ under the integrals.
	
Before proceeding with the analysis, we would like to introduce the notion of return time and comment on the singularity structure of $\F[][(1)]$. Consider any right-moving component of $\psi$ that intersects the detector at some proper time $\tau$. When propagated backwards along the null line $u(\tau)$, the right-mover will be reflected by the wall at $v = w(u(\tau))$ and becomes a left-mover which may intersect the detector at some earlier proper time $\tau_e$. This is guaranteed if the detector follows a trajectory which is asymptotically inertial in the far past. In such cases, we define the return time $\Delta{t}_{\mathrm{ret}}(\tau)$ at proper time $\tau$ as the difference $\tau - \tau_{e}$. To determine the quantity, notice that if we start from $\tau_{e}$ and propagate the left-mover forward in time, it will intersect the wall at $v = v(\tau_{e})$. Hence, the return time $\Delta{t}_{\mathrm{ret}}(\tau)$ satisfies the equation
\begin{align}
v(\tau - \Delta{t}_{\mathrm{ret}}(\tau)) = w(u(\tau))
	\implies \Delta{t}_{\mathrm{ret}}(\tau) = \tau - v^{-1}(w(u(\tau))).
\end{align}
Notice that $s = \Delta{t}_{\mathrm{ret}}(r)$ is precisely where the integrand of $\F[][(1)]$ is singular when $\epsilon = 0$. Non-existence of $\Delta{t}_{\mathrm{ret}}(r)$ for a given $r$ then implies that the first factor in the integrand is regular at that $r$. Note also that the condition $v(r) - w(u(r-s)) \geq v(r) - w(u(r)) \geq h_{\mathrm{min}}$, which is true for any non-intersecting timelike detector and wall trajectories, implies that the second factor of the integrand has no singular points for any $r \in \mathbb{R}$.

\subsection{Normally Ordered Coupling}

Instead of \eqref{Interaction Hamiltonian}, one may also consider the interaction Hamiltonian \cite{Hummer:2015xaa}
\begin{align} \label{Normally Ordered Interaction Hamiltonian}
	H_{\mathrm{int}}^{\mathrm{no}} = c \mu(\tau) \chi(\tau) :\overline{\psi}(u(\tau),v(\tau)) \psi(u(\tau),v(\tau)):,
\end{align}
where $:\overline{\psi}\psi:$ denotes a normally-ordered scalar density. In this case, only the first term of \eqref{Two-point function in terms of propagators} contributes to the two-point function $W^{(2,\bar{2})}(u,v;u',v')$ \cite{Louko:2016ptn}. The divergence issue due to the second term of \eqref{Two-point function in terms of propagators} hence does not arise. Indeed, the normally-ordered interaction Hamiltonian \eqref{Normally Ordered Interaction Hamiltonian} is proposed in \cite{Hummer:2015xaa} precisely to address the divergence issue.

In Minkowski spacetime without a wall, the end result is the same whether one discards the divergence by means of operation ordering, as we did in Appendix \ref{Regularisation of Two-Point Function}, or by working with \eqref{Normally Ordered Interaction Hamiltonian}. However, in the presence of a boundary, this is not the case. A detector interacting via the non-normally-ordered Hamiltonian \eqref{Interaction Hamiltonian} picks up a finite contribution --- that is $\F[][(2)]$ in \eqref{Detector Response} --- from the second term of \eqref{Two-point function in terms of propagators}. The response function of a detector interacting via the normally-ordered Hamiltonian \eqref{Normally Ordered Interaction Hamiltonian}, on the other hand, consists only of $\F[][(0)]$ and $\F[][(1)]$. In what follows, we will focus primarily on a detector interacting via \eqref{Interaction Hamiltonian} and make occasional comments about one interacting via \eqref{Normally Ordered Interaction Hamiltonian}.

\subsection{The Transition Rate} \label{The Transition Rate}

We now specialise to a detector resting at $z=d$ and parametrise its worldline  as
\begin{align} \label{Detector at Rest Trajectory}
	(u(\tau),v(\tau)) = (\tau, \tau + 2d).
\end{align}
The no-collision condition now reads $\tau + 2d - w(\tau) \geq h_{\mathrm{min}}$. Since $\Delta{t}_{\mathrm{ret}}(\tau) = \tau + 2d - w(\tau)$ in this case, we incidentally have $\Delta{t}_{\mathrm{ret}}(\tau) \geq h_{\mathrm{min}}$ .

Using methods in \cite{Hodgkinson:2013tsa,Satz:2006kb,Louko:2007mu,Hodgkinson:2011pc}, we may then evaluate the limit $\epsilon \to 0_{+}$ of $\F[][(0)]$ explicitly and obtain
\begin{align}
	\F[][(0)]
		&= - \frac{\omega}{2\pi} \int_{-\infty}^{\infty} \dd{r} [\chi(r)]^{2}
			+ \frac{1}{\pi^{2}} \int_{0}^{\infty} \frac{\dd{s}}{s^{2}}
				\int_{-\infty}^{\infty} \dd{r} \chi(r) [\chi(r) - \chi(r-s)]
					\nonumber \\
		&\qquad
			+ \frac{1}{\pi^{2}} \int_{-\infty}^{\infty} \dd{r} \chi(r)
				\int_{0}^{\infty} \dd{s} \chi(r-s) \left( \frac{1}{s^2}
			 - \frac{\cos(\omega s)\sqrt{w'(r) w'(r-s)}}{s(w(r) - w(r-s))} \right).
\end{align}
For $\F[][(1)]$, setting $\epsilon = 0$ in the second factor introduces an error of order $\epsilon$ which hence vanishes as we take the limit $\epsilon \to 0_{+}$. The remaining term, viewed as an integral over $r$, has an integrand that is bounded uniformly by an $\epsilon$-independent constant. Together with the fact that $\chi$ has a compact support, we may use the dominated convergence theorem to justify commuting the limit $\epsilon \to 0_{+}$ through the outer $r$-integral and obtain
\begin{align} \label{FMW1 of Detector at Rest}
	\F[][(1)]
		&= \frac{1}{\pi^{2}} \Re{ \int_{-\infty}^{\infty} \dd{r} \chi(r)
			\lim_{\epsilon \to 0_{+}} \int_{0}^{\infty} \dd{s}
			\frac{e^{- i \omega s} \chi(r-s) \sqrt{w'(r) w'(r-s)}}
				{(s - \Delta{t}_{\mathrm{ret}}(r) - i \epsilon)(s + \Delta{t}_{\mathrm{ret}}(r-s))} }.
\end{align}
One may then proceed to evaluate the limit $\epsilon \to 0_{+}$ using the Sokhotsky formula \eqref{Sokhotsky formula} if the singularity $s = \Delta{t}_{\mathrm{ret}}(r)$ is within the support of $\chi(r-s)$. Otherwise, if the singularity is outside the support or on the boundary, one may simply set $\epsilon = 0$ under the integral. The latter is justified by the fact that the singularity is suppressed by the switching function. This can be seen by evaluating the inner $s$-integral for arbitrary $\chi$ using the Sokhotsky formula, then specialising to one where $\chi(r-s) = 0$ for $s \geq \Delta{t}_{\mathrm{ret}}(r)$. For our purpose here, $\F[][(1)]$ as given by \eqref{FMW1 of Detector at Rest} is sufficient. As for $\F[][(2)]$, a substitution of \eqref{Detector at Rest Trajectory} into \eqref{FMW2} gives
\begin{align}
	\F[][(2)]
		&= \frac{2}{\pi^{2}} \int_{-\infty}^{\infty} \dd{r} \chi(r) \frac{\sqrt{w'(r)}}{\Delta{t}_{\mathrm{ret}}(r)}
			\int_{0}^{\infty} \dd{s} \chi(r-s) \cos(\omega s) \frac{\sqrt{w'(r-s)}}{\Delta{t}_{\mathrm{ret}}(r-s)}.
\end{align}

To analyse time-dependent situations, we are interested in how the transition probability of the detector changes over time. To this aim, following \cite{Satz:2006kb}, we consider the instantaneous transition rate of the detector that is defined as follows. Suppose that the switching function takes the form
\begin{align} \label{Family of switching functions}
	\chi(r) = h_{1} \left( \frac{r - \tau_{0} + \delta}{\delta} \right)
		\times h_{2} \left( \frac{- r + \tau + \delta}{\delta} \right),
\end{align}
where $\tau$ and $\tau_{0}$ are real parameters satisfying $\tau > \tau_{0}$, $\delta$ is a small positive parameter, and $h_{1}$ and $h_{2}$ are smooth non-negative functions such that $h_{1}(x) = h_{2}(x) = 0$ for $x \leq 0$ and $h_{1}(x) = h_{2}(x) = 1$ for $x \geq 1$. Then, the detector-field interaction is smoothly switched on during the interval $(\tau_{0}-\delta,\tau_{0})$ according to the function $h_{1}$, stays at a constant coupling strength $c$ for the interval $\Delta\tau = \tau - \tau_{0}$ and is smoothly switched off during the interval $(\tau,\tau+\delta)$ according to the function $h_{2}$.

The instantaneous transition rate of the detector is then defined as $\R := \partial_{\tau} \F$. Operationally, $\R$ cannot be measured by a single or even an ensemble of particle detectors since a measurement would change the initial condition for the subsequent dynamics of the detector. Instead, we need an ensemble of ensembles of particle detectors to measure $\R$ \cite{Louko:2006zv,Langlois:2005if}. Nevertheless, $\R$ has a very useful intuitive physical interpretation. It measures how the transition probability would change if the detector continues to interact with the field, at a constant coupling $c$, for an infinitesimal extra time. Taking the limit $\delta \to 0$ to eliminate any switching effect, we find that the transition rate of our detector is given by
\begin{align}
	\R = \R[][(0)] + \R[][(1)] + \R[][(2)],
		\label{RMW Detector at Rest}
\end{align}
where
\begin{align}
	\R[][(0)]
		&= - \frac{\omega}{\pi} \Theta(-\omega)
			+ \frac{1}{\pi^{2}} \int_{0}^{\Delta\tau} \dd{s} \cos(\omega s)
				\left( \frac{1}{s^{2}} - \frac{\sqrt{w'(\tau) w'(\tau-s)}}{s(w(\tau) - w(\tau-s))} \right)
					\nonumber \\
		&\qquad
			+ \frac{1}{\pi^{2}} \int_{\Delta\tau}^{\infty} \dd{s} \frac{\cos(\omega s)}{s^{2}},
				\label{RMW0 Detector at Rest} \\
	\R[][(1)]
		&= \lim_{\epsilon \to 0_{+}} \frac{1}{\pi^{2}} \int_{0}^{\Delta\tau} \dd{s}
			\Re{ \frac{e^{- i \omega s} \sqrt{w'(\tau) w'(\tau-s)}}{(s - \Delta{t}_{\mathrm{ret}}(\tau) - i \epsilon)(s + \Delta{t}_{\mathrm{ret}}(\tau-s))} },
				\label{RMW1 Detector at Rest} \\
	\R[][(2)]
		&= \frac{2 \sqrt{w'(\tau)}}{\pi^{2} \Delta{t}_{\mathrm{ret}}(\tau)}
			\int_{0}^{\Delta\tau} \dd{s} \cos(\omega s)
				\frac{\sqrt{w'(\tau-s)}}{\Delta{t}_{\mathrm{ret}}(\tau-s)}.
				\label{RMW2 Detector at Rest}
\end{align}

A few technical remarks are in order. First, similar to \eqref{FMW1 of Detector at Rest}, the evaluation of $\epsilon \to 0_{+}$ limit in \eqref{RMW1 Detector at Rest} depends on the relationship between $\Delta\tau$ and $\Delta{t}_{\mathrm{ret}}(\tau)$. We may simply set $\epsilon = 0$ under the integral when $\Delta\tau < \Delta{t}_{\mathrm{ret}}(\tau)$ and use the Sokhotsky formula \eqref{Sokhotsky formula} when $\Delta\tau > \Delta{t}_{\mathrm{ret}}(\tau)$. However, when $\Delta\tau = \Delta{t}_{\mathrm{ret}}(\tau)$, the limit is undefined since the singularity occurs at the boundary of integration interval and there is no switching function in \eqref{RMW1 Detector at Rest} to suppress the divergence. Hence, the expressions above are valid only when $\Delta\tau \neq \Delta{t}_{\mathrm{ret}}(\tau)$. Second, the integrand of $\R[][(2)]$ oscillates about a non-zero constant at large $s$. We thus have to keep the switch-on time $\tau_{0}$ strictly finite. Third, if one is working with the normally-ordered interaction Hamiltonian \eqref{Normally Ordered Interaction Hamiltonian}, $\R[][(2)]$ does not contribute to the transition rate and we may push the switch-on time $\tau_{0}$ to the asymptotic past. When this is done, the last term in \eqref{RMW0 Detector at Rest} is $\BigO(1/\Delta\tau)$ and the limit $\epsilon \to 0_{+}$ in \eqref{RMW1 Detector at Rest} should only be evaluated using the Sokhotsky formula \eqref{Sokhotsky formula}.

Let us now compare the transition rate \eqref{RMW Detector at Rest} of our detector to that of a detector interacting with a scalar field \cite{Hodgkinson:2013tsa} or its derivative \cite{Juarez-Aubry:2014jba} in a moving mirror spacetime. In the case of a detector coupled linearly to a scalar field, the corresponding transition rate of the detector (that is also at rest with its trajectory parametrised by \eqref{Detector at Rest Trajectory}) is given by
\begin{align} \label{Scalar detector linear transition rate}
	\R[\phi]
		&= - \frac{1}{2\pi} \int_{0}^{\Delta\tau} \dd{s} \cos(\omega s)
			\ln(\frac{(w(\tau)-w(\tau-s))s}{(s + \Delta{t}_{\mathrm{ret}}(\tau-s))
				\abs{s - \Delta{t}_{\mathrm{ret}}(\tau)}}) \nonumber \\
		&\qquad
			- \frac{1}{2} \int_{0}^{X} \dd{s} \sin(\omega s),
\end{align}
where $X = \Delta\tau$ or $\Delta{t}_{\mathrm{ret}}(\tau)$, whichever shorter at the detector's proper time $\tau$. Note that $\R[\phi]$ as given in \eqref{Scalar detector linear transition rate} above is not exactly identical to that obtained in \cite{Hodgkinson:2013tsa}. This is because, while we have kept the switch-on time finite for the purpose of comparison, Hodgkinson pushed the switch-on time to the far past in \cite{Hodgkinson:2013tsa}. To circumvent the resultant infinity that arose, he employed a different family of switching functions than \eqref{Family of switching functions}. Nevertheless, the main feature that we would like to highlight here, that is, the integrand of $\R[\phi]$ is a logarithmic function of $w(\tau)$, is similar. Meanwhile, in the case of a detector coupled linearly to the derivative of a scalar field, accounting for the slight difference in parametrising the detector's trajectory and the finite switch-on time in our analysis, the corresponding transition rate is given by \cite{Juarez-Aubry:2014jba}
\begin{align} \label{Scalar detector derivative transition rate}
	\R[\partial\phi] = \R[\partial\phi][(0)] + \R[\partial\phi][(1)]
\end{align}
where
\begin{align}
	\R[\partial\phi][(0)]
		&= - \omega \Theta(-\omega)
			+ \frac{1}{2\pi} \int_{0}^{\Delta\tau} \dd{s} \cos(\omega s)
				\left( \frac{1}{s^{2}} - \frac{w'(\tau) w'(\tau-s)}{(w(\tau) - w(\tau-s))^{2}} \right)
					\nonumber \\
		&\qquad
			+ \frac{1}{\pi} \int_{\Delta\tau}^{\infty} \dd{s} \frac{\cos(\omega s)}{s^{2}}, \\
	\R[\partial\phi][(1)]
		&= \frac{1}{2\pi} \int_{0}^{\Delta\tau} \dd{s}
			\Re{ e^{- i \omega s} \frac{w'(\tau-s)}{(s + \Delta{t}_{\mathrm{ret}}(\tau-s))^{2} } }
				\nonumber \\
		&\qquad
			+ \lim_{\epsilon \to 0_{+}} \frac{1}{2\pi} \int_{0}^{\Delta\tau} \dd{s}
			\Re{ e^{- i \omega s} \frac{w'(\tau)}{(s - \Delta{t}_{\mathrm{ret}}(\tau) - i \epsilon)^{2} } }.
\end{align}
Comparing \eqref{RMW Detector at Rest} to \eqref{Scalar detector linear transition rate} and \eqref{Scalar detector derivative transition rate}, we clearly see that the transition rate \eqref{RMW Detector at Rest} of our detector takes a different form. Except possibly for some very specific wall/mirror trajectories, we can expect that $\R[][(0)]$, $\R[\phi][(0)]$ and $\R[\partial\phi][(0)]$ to be different from each other at any given detector proper time $\tau$. Hence, unlike an observer who only analyses the stress-energy tensor of the field, an observer equipped with an Unruh-DeWitt detector will generically be able to distinguish a Dirac fermion from a scalar boson. In the next section, we will see an example where this is indeed the case.

\subsection{Late Time Thermal Radiation Model} \label{Section Collapsing Star Model}

Of particular interest to us is the detector transition rate at late times when the wall follows a trajectory that produces Hawking-like radiation to infinity in the far future. Such a trajectory has the characteristic asymptotic behaviour \cite{Davies:1977yv,Birrell:1982ix}
\begin{align} \label{Collapsing Star Trajectory Asymptotic Form}
	z_{\mathrm{w}}(t) \to - t - A e^{-2at} + B \qquad \text{as } t \to \infty,
\end{align}
where $a$, $A$ and $B$ are positive constants. The parameter $a$, in particular, plays the role of the surface gravity $\kappa$ in the Bekenstein-Hawking temperature \eqref{Bekenstein-Hawking temperature}. For concreteness, let us consider the trajectory \cite{Hotta:1994ha,Good:2011,Hodgkinson:2013tsa,Juarez-Aubry:2014jba,Good:2013lca,Good:2017ddq}
\begin{align} \label{Collasing Star Trajectory 1}
	w(u) = - \frac{1}{a} \ln(1 + e^{-au}),
\end{align}
where $a > 0$. Using the formula $\sinh^{-1}(x) = \ln(x + \sqrt{x^{2} + 1})$, we obtain that
\begin{align} \label{Collasing Star Trajectory 2}
	z_{\mathrm{w}}(t)
		= \frac{1}{a} \ln\left( - \frac{e^{at}}{2} + \sqrt{\frac{e^{2at}}{4} + 1} \right).
\end{align}
At early times, the wall moves with an inertial motion as required. The trajectory \eqref{Collasing Star Trajectory 1} asymptotes to $z_{\mathrm{w}}(t) = 0$ from the left as $t \to -\infty$. At late times, the wall approaches a null trajectory with $v = 0$ as the asymptote and \eqref{Collasing Star Trajectory 2} satisfies \eqref{Collapsing Star Trajectory Asymptotic Form} with $A = 1$ and $B = 0$.

Two remarks are in order. First, only the late time limit of the trajectory \eqref{Collasing Star Trajectory 1}, or equivalently \eqref{Collasing Star Trajectory 2}, is relevant in modelling a collapsing star. Even then, the late time limit of \eqref{Collasing Star Trajectory 2} only models the late time phase of the collapsing star. For a wall trajectory that exactly mimics a collapsing null shell at all times, we refer the readers to \cite{Good:2016oey}. Second, the proper acceleration $\alpha_{\mathrm{w}}$ of a wall following the trajectory \eqref{Collasing Star Trajectory 2} is not constant. In particular, $\alpha_{\mathrm{w}}$ is not equal to the parameter $a$ albeit the latter's role as the surface gravity when modelling the late time phase of a collapsing star. Instead, the wall's proper acceleration is given by $\alpha_{\mathrm{w}}(t) = -(a/2) \exp(at)$, which diverges at late times. In terms of the wall's proper time $\tau_{\mathrm{w}}$, where the late time limit is given by $\tau_{\mathrm{w}} \to 0_{-}$, the proper acceleration approaches the characteristic scale-independent acceleration $\alpha_{\mathrm{w}}(\tau_{\mathrm{w}}) = 1/\tau_{\mathrm{w}}$ that produces a constant flux of thermal radiation \cite{Carlitz:1986nh,Good:2017ddq}. 

From previous particle detector calculations involving a massless scalar field \cite{Hodgkinson:2013tsa,Juarez-Aubry:2014jba}, one may expect that the late time transition rate in our case here to contain a term proportional to the Fermi-Dirac statistics, that is
\begin{align} \label{Expected late time transition rate}
	\R
		&= f(\omega,a) \Theta(-\omega) + \frac{g(\omega,a)}{1 + e^{2\pi\omega/a}} + h(\omega,a),
\end{align}
where $f$, $g$ and $h$ are some functions of $\omega$ and $a$. The first and second terms represent the contributions from undisturbed left-movers and thermal right-movers respectively. However, as shown in Appendix \ref{Late Times Transition Rate}, this is not the case. The late time transition rate of a static detector at $z=d$, interacting with the field via either interaction Hamiltonian \eqref{Interaction Hamiltonian} or \eqref{Normally Ordered Interaction Hamiltonian}, is in fact given by
\begin{align} \label{Late time transition rate for collapse model}
	\R
		&= \frac{a}{2\pi^{2}} \ln(1 + e^{-\frac{2\pi\omega}{a}}) + \smallO(1)
\end{align}
when it is switched on at an arbitrary fixed finite $\tau_{0}$. The same holds when the detector interacts via \eqref{Normally Ordered Interaction Hamiltonian} and is switched on in the asymptotic past. Comparing \eqref{Late time transition rate for collapse model} to \eqref{Expected late time transition rate}, two observations can be made.

First, the separation between contributions from undisturbed left-movers and thermal right-movers is not apparent in \eqref{Late time transition rate for collapse model}. This can actually be traced back to the quadratic nature of the detector-field coupling whose general feature is to couple the left-moving and right-moving components of the field in $W^{(2,\bar{2})}$. On hindsight, due to this coupling between left-movers and right-movers, the form \eqref{Expected late time transition rate} should not be expected to hold.

Second, \eqref{Late time transition rate for collapse model} contains no term that is proportional to the number density of fermions with energy $\omega$ in a fermionic thermal bath at temperature $a/2\pi$, as suggested in \eqref{Expected late time transition rate}. Our detector hence is not counting the number of fermions with energy $\omega$ in usual sense of particle counting. Recall, nevertheless, that the Helmholtz free energy of fermions in length $L$ at temperature $T = a/2\pi$ is given by \cite{Kapusta:2006pm}
\begin{align}
	F = - 2 L \int_{0}^{\infty} \frac{\dd{\omega}}{2\pi} \left[ \omega
			+ 2 \times \frac{a}{2\pi} \ln(1 + e^{-\frac{2\pi\omega}{a}}) \right].
\end{align}
The first term is the vacuum energy, hence may be ignored. The factor two in the second term accounts for the existence of fermions and anti-fermions. Meanwhile, the overall factor two outside the integral accounts for the existence of a left-mover and a right-mover for each energy. We see that, for $\omega > 0$, the transition rate \eqref{Late time transition rate for collapse model} is actually proportional to the Helmholtz free energy density of fermions with energy $\omega$ in a thermal bath at temperature $a/2\pi$. An observer equipped with a detector will therefore be able to infer that the late time radiation from the wall is made of fermions.

\section{Discussion} \label{Discussion}

In this paper, we have analysed the moving wall model in $1+1$ Minkowski spacetime for a massless spinor field governed by the Dirac equation. To ensure a well-defined early time state, we have restricted ourselves to a wall that follows an inertial trajectory in the far past. When the field is in a state corresponding to early time vacuum, we found that the wall radiates in such a way that the stress-energy tensor \eqref{Renormalised Stress-Energy Tensor Result} of the spinor field is exactly equal to that of a massless scalar field at all times \cite{Davies:1977yv,Birrell:1982ix,Good:2011}. This implies that an observer examining only the stress-energy tensor is unable to tell whether the radiation is made of Dirac fermions or scalar bosons.

We then considered an observer who analyses the radiation using an Unruh-DeWitt detector coupled linearly to the scalar density of the spinor field. Specifically, we focused on a static observer and calculated the transition rate of the detector. Even though it is operationally difficult to be measured since an observer requires an ensemble of ensembles of detectors to do so, the transition rate has an intuitive physical interpretation. It quantifies how the transition probability of the detector would change when the detector continues to interact with the field at a constant coupling strength for an infinitesimal additional time. We found that the transition rate is finite at almost all times. It is divergent only when the detector-field interaction period coincides with the time taken for the field to be reflected back, backwards in time, to the observer. When compared to the corresponding result for a massless scalar field \cite{Hodgkinson:2013tsa,Juarez-Aubry:2014jba}, we found that the transition rate \eqref{RMW Detector at Rest} does indeed take a different form. Hence, an observer equipped with an Unruh-DeWitt detector will generically be able to distinguish a Dirac fermion from a scalar boson.

As an illustration, we then considered a wall trajectory for which a mirror reflecting a scalar field is known to emit thermal radiation in the far future \cite{Hodgkinson:2013tsa,Juarez-Aubry:2014jba}. For the wall reflecting a fermion field, we obtained an interesting result for the detector's late time transition rate; the detector clicks at a rate proportional to the Helmholtz free energy density of  fermions, whose energy matches the detector energy gap $\omega$, in a thermal bath at temperature $T=a/2\pi$. On one hand, this is to be contrasted with the expectation that the detector counts the number of particles, hence should contain a term proportional to the number density of thermal fermions with energy $\omega$. On the other hand, the late time transition rate is to be compared with the corresponding result for a massless scalar field $\phi$ where the detector partially clicks at a rate proportional to the number density of bosons, with energy $\omega$, in a thermal bath of the same temperature \cite{Hodgkinson:2013tsa,Juarez-Aubry:2014jba}. The latter comparison shows that our second observer can indeed conclude that the radiation consists of fermions.

There are several limitations in our analysis that could serve as a source of ideas for future projects. First, we have focused exclusively on the response of a static detector. It would be interesting to see how the detector's motion, inertial or otherwise, would change the result. Second, the scalar detectors that our results are compared against couple linearly either to the scalar field itself or its derivative. It would be interesting to consider a scalar detector that couples linearly to the scalar density $\phi^{\dagger}\phi$ of the scalar field. In particular, we see in Section \ref{Section Collapsing Star Model} that the left-moving and right-moving components of the spinor field conspire in such a way that the late time transition rate of our detector is proportional to the Helmholtz free energy density of mode $\omega$ in a fermionic thermal bath. It would be interesting to see if a similar phenomenon occurs for the $\phi^{\dagger}\phi$ detector. Third, only a wall trajectory where the late time limit satisfies the late time thermality condition of a collapsing star has been considered in detail. Another interesting wall trajectory is one where the wall accelerates uniformly for a certain period of time. The similarity between the stress-energy tensor of a scalar field and a spinor field implies that the wall analysed in this paper also does not radiate when it is accelerating uniformly. For a scalar field, Davies and Fulling argued that a detector detects particle fluxes albeit the absence of any radiation \cite{Davies:1977yv}. However, Groove later made an analysis that resolves the discrepancy \cite{Grove:1986fy}. After identifying an approximation that led Davies and Fulling to their conclusion, Groove showed that a detector operating only in the region where the mirror accelerates uniformly indeed detects no particle. A calculation of the transition rate for our detector in a similar setting would thus complement Groove's work.

\section*{Acknowledgements}
The author thanks Jorma Louko for helpful discussions and the anonymous referees for their constructive comments. The author is financially supported by the Ministry of Higher Education, Malaysia and Universiti Sains Malaysia.

\appendix
\section{Regularised Two-Point Function} \label{Regularisation of Two-Point Function}

In this appendix, we note the technical issues in interpreting $W^{(2,\bar{2})}$ as a distribution and outline how \eqref{Regularised Two-Point Function} is obtained.

Let us start with the first term in \eqref{Two-point function in terms of propagators}. Viewed as distributions, $S^{+}$ and $S^{-}$ are well-defined individually as the $\epsilon \to 0_{+}$ limits of \eqref{Splus Distribution} and \eqref{Sminus Distribution} respectively. Upon multiplication, each factor should \textit{a priori} comes with its own $\epsilon$ parameter, say $\epsilon_{+}$ and $\epsilon_{-}$, that need not be equal or even related to each other. This gives us
\begin{align} \label{TrSplusSminus with distinct epsilons}
	&\Tr{S^{+}(u,v;u',v')S^{-}(u',v';u,v)} \nonumber \\
		&\quad =
			- \frac{1}{4\pi^{2}} \bigg[ \frac{ \sqrt{w'(u)w'(u')} }{ ( w(u) - w(u') - i \epsilon_{+} ) ( v - v' - i \epsilon_{-} ) }
			+ \frac{ \sqrt{w'(u)w'(u')} }{ ( w(u) - w(u') - i \epsilon_{-} ) ( v - v' - i \epsilon_{+} ) }
				\nonumber \\
		&\quad \quad
			- \frac{ \sqrt{w'(u)w'(u')} }{ ( w(u) - v' - i \epsilon_{+} ) ( v - w(u') - i \epsilon_{-} ) }
			- \frac{ \sqrt{w'(u)w'(u')} }{ ( w(u) - v' - i \epsilon_{-} ) ( v - w(u') - i \epsilon_{+} ) } \bigg],
\end{align}
where the limits $\epsilon_{\pm} \to 0_{+}$ are implied. When pulled back onto the detector's worldline to evaluate the corresponding detector response contribution, the order in which the limits are taken could be expected to matter since such limits need not, in general, commute. However, notice that each term in \eqref{TrSplusSminus with distinct epsilons} is a product of two distributions with different sets of variables. Since each factor in each term may be evaluated, after a suitable change of variables, using the Sokhotsky formula
\begin{align} \label{Sokhotsky formula}
	\lim_{\epsilon \to 0_{+}} \frac{1}{x - i \epsilon}
		= \pv \left( \frac{1}{x} \right) + i \pi \delta(x),
\end{align}
where $\pv(1/x)$ denotes the Cauchy principal value, \eqref{TrSplusSminus with distinct epsilons} does define the same well-behaved distribution on $\mathbb{R}^{4}$ regardless of the order in which the two limits are taken. Hence, we may set $\epsilon_{+} = \epsilon_{-}$ and obtain the first two terms of \eqref{Regularised Two-Point Function} before pulling them back onto the detector's worldline. We note here that this way of defining a product of two distributions is implicit in \cite{Louko:2016ptn}.

Now consider the second term in \eqref{Two-point function in terms of propagators}. Due to the coincidence limit, $S^{-}$ in this term is not well-defined \cite{Louko:2016ptn}. However, we see from \eqref{Sminus Distribution} that the traces of the ill-defined terms, that is the first and last term of \eqref{Sminus Distribution}, are proportional to $U_{+}^{\dagger} \beta U_{+}$ and $U_{-}^{\dagger} \beta U_{-}$ respectively, both of which vanish by virtue of \eqref{Action of Dirac matrices on U basis} and the orthogonality of the spinor basis. Hence, if we take the spinorial trace in $\Tr{S^{-}(u,v;u,v)}$ before performing any other operations, that is by defining $\Tr{S^{-}(u,v;u,v)}$ as the coincidence limit of $\Tr{S^{-}(u,v;u',v')}$, then each factor in the second term of \eqref{Two-point function in terms of propagators} becomes a well-defined distribution. Furthermore, since each factor is defined on different $\mathbb{R}^{2}$, the product is also well-defined regardless of whether the two $\epsilon$ parameters associated to each factor are distinct or otherwise. Accepting this definition of $\Tr{S^{-}(u,v;u,v)}$ and choosing to work with a single $\epsilon$, we obtain 
\begin{align}
	\Tr{S^{-}(u,v;u,v)}\Tr{S^{-}(u',v';u',v')}
		\!=\! \frac{1}{\pi^{2}} \frac{ \sqrt{w'(u)} (w(u) - v) }{ ((w(u) - v)^{2} + \epsilon^{2}) }
			\frac{ \sqrt{w'(u')} (w(u') - v') }{ ((w(u') - v')^{2} + \epsilon^{2}) },
\end{align}
which is the last term of \eqref{Regularised Two-Point Function}.

\section{Late Times Transition Rate} \label{Late Times Transition Rate}

In this appendix, we derive the late time transition rate \eqref{Late time transition rate for collapse model}.

\subsection{Finite Switch-On Time ($H_{\mathrm{int}}$ or $H_{\mathrm{int}}^{\mathrm{no}}$)}

We start by considering a detector that is switched on at a strictly finite $\tau_{0}$. The detector may interact either via \eqref{Interaction Hamiltonian} or \eqref{Normally Ordered Interaction Hamiltonian}. For the latter, ignore the analysis involving $\R[][(2)]$.

Consider first $\R[][(0)]$. Substituting the trajectory \eqref{Collasing Star Trajectory 1} into \eqref{RMW0 Detector at Rest} and letting $h(\tau) = (1 + e^{a\tau})^{-1}$, we have
\begin{align} \label{RMW0 Collapse Trajectory 0}
	\R[][(0)]
		&= - \frac{\omega}{\pi} \Theta(-\omega)
			+ \frac{1}{\pi^{2}} \int_{0}^{\Delta\tau} \dd{s} \cos(\omega s)
				\left( \frac{1}{s^{2}} - X_{\tau}(s) \right)
					\nonumber \\
		&\qquad
			+ \frac{1}{\pi^{2}} \int_{\Delta\tau}^{\infty} \dd{s} \frac{\cos(\omega s)}{s^{2}},
\end{align}
where 
\begin{align}
	X_{\tau}(s)
		= \frac{ah(\tau)e^{\frac{as}{2}}}
			{s(1 + h(\tau)(e^{as}-1))^{\frac{1}{2}}\ln(1 + h(\tau)(e^{as}-1))}.
\end{align}
Adding and subtracting an integral that has an integrand identical to the second term of \eqref{RMW0 Collapse Trajectory 0}, including the factor $\pi^{-2}$, but integrates from $s = \Delta\tau$ to $s = \infty$, we obtain
\begin{align} \label{RMW0 Collapse Trajectory 1}
	\R[][(0)]
		&= - \frac{\omega}{\pi} \Theta(-\omega)
			+ \frac{1}{\pi^{2}} \int_{0}^{\infty} \dd{s} \cos(\omega s)
				\left( \frac{1}{s^{2}} - X_{\tau}(s) \right)
					\nonumber \\
		&\qquad
			+ \frac{1}{\pi^{2}} \int_{\Delta\tau}^{\infty} \dd{s} \cos(\omega s) X_{\tau}(s).
\end{align}

Focusing on $s > 0$ that we are concerned with, partially differentiating $X_{\tau}(s)$ with respect to $\tau$ or $h$ shows that $X_{\tau}(s)$ for fixed $s$ monotonically decreases as $\tau$ increases. The upper and lower bounds of $X_{\tau}(s)$ are thus given by $X_{-}(s)$ and $X_{+}(s)$ respectively where
\begin{align}
	X_{-}(s) &= \lim_{\tau \to -\infty} X_{\tau}(s) = \frac{1}{s^{2}},
		\label{Xminus} \\
	X_{+}(s) &= \ \lim_{\tau \to \infty} X_{\tau}(s) \ = \frac{a}{2s\sinh(\frac{as}{2})}.
		\label{Xplus}
\end{align}

Consider the limit $\tau \to \infty$ of $\R[][(0)]$. From the upper bound of $X_{\tau}(s)$, which is given by \eqref{Xminus}, it follows that the last term in \eqref{RMW0 Collapse Trajectory 1} is $\BigO(1/\Delta\tau)$. Adding and subtracting $\cos(\omega s) X_{+}(s)$ under the first integral in \eqref{RMW0 Collapse Trajectory 1}, and using the fact that $\cos(\omega s)[X_{+}(s) - 1/s^2]$ is an even function of $s$, we have
\begin{align} \label{RMW0 Collapse Trajectory 2}
	\R[][(0)]
		&= - \frac{\omega}{\pi} \Theta(-\omega)
			+ \frac{1}{2\pi^{2}} \int_{-\infty}^{\infty} \dd{s} \cos(\omega s)
				\left( \frac{1}{s^{2}} - X_{+}(s) \right)
					\nonumber \\
		&\qquad
			- \frac{1}{\pi^{2}} \int_{0}^{\infty} \dd{s} \cos(\omega s)
				\left( X_{\tau}(s) - X_{+}(s) \right)
			+ \BigO(\frac{1}{\Delta\tau}).
\end{align}
Since $X_{\tau}(s) - X_{+}(s) \to 0$ pointwise in a monotone manner for $s \geq 0$, the magnitude of the third term in \eqref{RMW0 Collapse Trajectory 2} decreases as $\tau$ increases and eventually vanishes in the limit $\tau \to \infty$. 

Deforming the integration contour of the second term in \eqref{RMW0 Collapse Trajectory 2} to a contour $C$ in the complex $s$ plane along the real axis but with a dip into the lower half-plane near $s=0$,
\begin{align} \label{RMW0 Collapse Trajectory 3}
	\R[][(0)]
		&= - \frac{\omega}{\pi} \Theta(-\omega)
			+ \frac{1}{2\pi^{2}} \int_{C} \dd{s} \frac{\cos(\omega s)}{s^{2}}
			- \frac{a}{4\pi^{2}} \int_{C} \dd{s} \frac{\cos(\omega s)}{s \sinh(\frac{as}{2})}
			+ \smallO(1).
\end{align}
Writing $\cos(\omega s) = 1 - (1 - \cos(\omega s))$, the integral factor of the second term in \eqref{RMW0 Collapse Trajectory 3} becomes a sum of two integrals, one with the integrand $s^{-2}$ and while the other $(1 - \cos(\omega s))s^{-2}$. Setting the dip to be an anticlockwise semicircle with radius $r > 0$, we find that the former vanishes. To calculate the latter, deform $C$ back to being a contour on the real line, integrate the integral by parts and use the identity $\int_{0}^{\infty} \sin(ax)/x \dd{x} = (\pi/2) \sgn(a)$. Collecting the results, we find that the first two terms of \eqref{RMW0 Collapse Trajectory 3} combine to give $\omega/2\pi$.

To evaluate the third term of \eqref{RMW0 Collapse Trajectory 3}, consider the integral factor
\begin{align} \label{I for RMW0 Collapse Trajectory 1}
	I(\omega) = \int_{C} \dd{s} \frac{\cos(\omega s)}{s \sinh(\frac{as}{2})}.
\end{align}
While the integrand of $I(\omega)$ has a singularity at $s=0$, the integrand of $\partial{I}/\partial\omega$, obtained by differentiating $I(\omega)$ with respect to $\omega$ under the integral, is regular on the real line. Deforming $C$ in $\partial{I}/\partial\omega$ back to being a contour on the real line and using 3.981.1 of \cite{Zwillinger2015}, we find that $\partial{I}/\partial\omega = - (2\pi/a) \tanh(\pi\omega/a)$. By integrating $\partial{I}/\partial\omega$ with respect to $\omega$, it follows that
\begin{align} \label{I for RMW0 Collapse Trajectory 2}
	I(\omega) = - 2 \ln(\cosh(\frac{\pi\omega}{a})) + I_{0},
\end{align}
where $I_{0}$ is independent of $\omega$. To determine the value of $I_{0}$, notice that it is the value of $I(\omega)$ when $\omega = 0$. Hence, consider $I(0)$ and deform $C$ so that it is the large $R$ limit of $C_{1} + C_{2} + C_{3}$ where $C_{1}$ runs from $s = - R$ to $s = - R - i\pi/a$ along $\Re{s} = - R$, $C_{2}$ runs from $s = - R - i\pi/a$ to $s = R - i\pi/a$ along $\Im{s} = - \pi/a$ and $C_{3}$ runs from $s = R - i\pi/a$ to $s = R$ along $\Re{s} = R$. The contributions from $C_{1}$ and $C_{3}$ are both $\BigO(1/(R\sinh(aR/2)))$ and hence vanish as $R \to \infty$. Parametrising $C_{2}$ as $x - i\pi/a$ for $x \in (-\infty,\infty)$ and using the identity $\sinh(x - i\pi/2) = - i \cosh(x)$, we find that
\begin{align} \label{I0 for RMW0 Collapse Trajectory}
	I(0) &= - \frac{a}{\pi} \int_{-\infty}^{\infty} \frac{\dd{x}}{( 1 + a^{2}x^{2}/\pi^{2} )( \cosh(ax/2) )}
			+ \frac{ia^{2}}{\pi^{2}} \int_{-\infty}^{\infty} \frac{x\dd{x}}{( 1 + a^{2}x^{2}/\pi^{2} )( \cosh(ax/2) )}.
\end{align}
The imaginary part of \eqref{I0 for RMW0 Collapse Trajectory} vanishes since the integrand is odd. The real part can be evaluated using 3.522.8 of \cite{Zwillinger2015} after a change of variable to $y = ax/\pi$, giving $I(0) = - 2 \ln 2$. Combining the results above, we find that, as $\tau \to \infty$,
\begin{align} \label{RMW0 Collapse Trajectory Final}
	\R[][(0)] = \frac{a}{2\pi^{2}} \ln(1 + e^{\frac{2\pi\omega}{a}}) + \smallO(1).
\end{align}

Let us now look at $\R[][(1)]$. To keep the analysis neat, we shall not write $w'(\tau)$ and $\Delta{t}_{\mathrm{ret}}(\tau)$ explicitly. For reference, we note here that
\begin{align}
	w'(\tau)
		&= \frac{1}{1 + e^{a\tau}},
			\label{Reference 1 for RMW1 late time limit} \\
	w''(\tau)
		&= - \frac{a w'(\tau)}{1 + e^{-a\tau}},
			\label{Reference 2 for RMW1 late time limit} \\
	\Delta{t}_{\mathrm{ret}}(\tau)
		&= \tau + 2d + \frac{1}{a} \ln(1 + e^{-a\tau})
			\label{Reference 3 for RMW1 late time limit}
\end{align}
for the trajectory \eqref{Collasing Star Trajectory 1}. In the case where $\tau_{0} \geq -2d$, we have $\Delta{t}_{\mathrm{ret}}(\tau) > \Delta\tau$ for all $\tau \in (\tau_{0},\infty)$. This implies that we may set $\epsilon = 0$ under the integral of \eqref{RMW1 Detector at Rest}. Since $\big| \cos(\omega s) \sqrt{w'(\tau) w'(\tau-s)} \big| \leq 1$ and $s + \Delta{t}_{\mathrm{ret}}(\tau-s) \geq \Delta\tau$ in this case, it follows that
\begin{align} \label{RMW1 Collapse Trajectory Late Times Upper Bound 1}
	\abs{\R[][(1)]}
		\leq - \frac{1}{\pi^{2}\Delta\tau} \ln(1-\frac{\Delta\tau}{\Delta{t}_{\mathrm{ret}}(\tau)}),
\end{align}
which is valid for any finite $\tau$. Implementing L'H\^{o}pital's rule to the right hand side of \eqref{RMW1 Collapse Trajectory Late Times Upper Bound 1}, it follows that $\R[][(1)] = \BigO(1/\Delta\tau)$ in the limit $\tau \to \infty$.

In the case of $-\infty < \tau_{0} < -2d$, there exists $\tau_{c}$ such that $\Delta\tau = \Delta{t}_{\mathrm{ret}}(\tau)$ when $\tau = \tau_{c}$ and $\Delta\tau > \Delta{t}_{\mathrm{ret}}(\tau)$ when $\tau > \tau_{c}$. Since we are interested in the late time limit $\tau \to \infty$, the latter will eventually necessarily be satisfied. Starting at some $\tau$ such that $\Delta\tau > \Delta{t}_{\mathrm{ret}}(\tau)$, we integrate \eqref{RMW1 Detector at Rest} by parts, integrating the factor $(s - \Delta{t}_{\mathrm{ret}}(\tau) - i \epsilon)^{-1}$ in the integrand, and take the limit $\epsilon \to 0_{+}$ to obtain
\begin{align} \label{RMW1}
	&\R[][(1)] \nonumber \\
		&\quad
			= \frac{1}{\pi^{2}} \bigg( \frac{\cos(\omega\Delta\tau)\sqrt{w'(\tau)w'(\tau_{0})}}{\Delta{t}_{\mathrm{ret}}(\tau) + w(\tau) - w(\tau_{0})}
				\ln(\Delta\tau - \Delta{t}_{\mathrm{ret}}(\tau)) - \frac{w'(\tau)}{\Delta{t}_{\mathrm{ret}}(\tau)} \ln(\Delta{t}_{\mathrm{ret}}(\tau)) \bigg)
					\nonumber \\
		&\quad \qquad
			+ \frac{1}{\pi^{2}} \sqrt{w'(\tau)} \bigg(
				\int_{0}^{\Delta\tau} \dd{s} \Re{q_{1}(\tau,s)} \ln|s-\Delta{t}_{\mathrm{ret}}(\tau)|
					\nonumber \\
		&\quad \qquad \qquad \qquad \qquad \qquad
			+ \frac{1}{2} \int_{0}^{\Delta\tau} \dd{s} \Re{q_{2}(\tau,s)} \ln|s-\Delta{t}_{\mathrm{ret}}(\tau)|
					\nonumber \\
		&\quad \qquad \qquad \qquad \qquad \qquad
			+ \omega \int_{0}^{\Delta\tau} \dd{s} \Re{q_{3}(\tau,s)}
				\ln|s-\Delta{t}_{\mathrm{ret}}(\tau)| \bigg)
					\nonumber \\
		&\quad \qquad
			- \frac{1}{\pi} \sqrt{w'(\tau)} \bigg(
				\int_{0}^{\Delta{t}_{\mathrm{ret}}(\tau)} \dd{s} \Re{iq_{1}(\tau,s)}
			+ \frac{1}{2} \int_{0}^{\Delta{t}_{\mathrm{ret}}(\tau)} \dd{s} \Re{iq_{2}(\tau,s)}
					\nonumber \\
		&\quad \qquad \qquad \qquad \qquad \qquad
			+ \omega \int_{0}^{\Delta{t}_{\mathrm{ret}}(\tau)} \dd{s} \Re{iq_{3}(\tau,s)}
				\bigg),
\end{align}
where
\begin{align}
	q_{1}(\tau,s)
		&= \frac{e^{-i \omega s} (w'(\tau-s))^{\frac{3}{2}}}
			{(s + \Delta{t}_{\mathrm{ret}}(\tau-s))^{2}}, \\
	q_{2}(\tau,s)
		&= \frac{e^{-i \omega s}}{s + \Delta{t}_{\mathrm{ret}}(\tau-s)} \frac{w''(\tau-s)}{\sqrt{w'(\tau-s)}}, \\
	q_{3}(\tau,s)
		&= \frac{i e^{-i \omega s}\sqrt{w'(\tau-s)}}{s + \Delta{t}_{\mathrm{ret}}(\tau-s)}.
\end{align}
The boundary terms in \eqref{RMW1}, that is the non-integral terms, clearly vanish as $\tau \to \infty$. For the integral terms, observe that, using \eqref{Reference 1 for RMW1 late time limit} -- \eqref{Reference 3 for RMW1 late time limit}, we have
\begin{align}
	\abs{q_{1}(\tau,s)} &= \abs{iq_{1}(\tau,s)}
		\leq \frac{1}{4d^{2}}, \\
	\abs{q_{2}(\tau,s)} &= \abs{iq_{2}(\tau,s)}
		\leq \frac{a}{2d}, \\
	\abs{q_{3}(\tau,s)} &= \abs{iq_{3}(\tau,s)}
		\leq \frac{1}{2d},
\end{align}
uniformly in $\tau$ for $s \geq 0$. Each of the first three integrals in \eqref{RMW1} then is bounded in magnitude by a $\tau$-independent constant times
\begin{align}
	J(\tau)
		&= \sqrt{w'(\tau)} \int_{0}^{\Delta\tau} \dd{s} |\ln|s-\Delta{t}_{\mathrm{ret}}(\tau)|| \nonumber \\
		&= \sqrt{w'(\tau)} \big[ 4 - \Delta\tau + \Delta{t}_{\mathrm{ret}}(\tau) \ln\Delta{t}_{\mathrm{ret}}(\tau) \nonumber \\
		&\qquad \qquad \qquad
			+ (\Delta\tau-\Delta{t}_{\mathrm{ret}}(\tau)) \ln(\Delta\tau-\Delta{t}_{\mathrm{ret}}(\tau)) \big],
\end{align}
while each of the last three integrals is bounded in magnitude by a $\tau$-independent constant times
\begin{align}
	K(\tau) = \sqrt{w'(\tau)} \int_{0}^{\Delta{t}_{\mathrm{ret}}(\tau)} \dd{s} = \sqrt{w'(\tau)} \Delta{t}_{\mathrm{ret}}(\tau).
\end{align}
As $\tau \to \infty$, both $J(\tau)$ and $K(\tau)$ are exponentially suppressed by the prefactor $\sqrt{w'(\tau)}$, implying that each integral term in \eqref{RMW1} vanishes at late times. Combining all the results above, it follows that $\R[][(1)]$ vanishes in the limit $\tau \to \infty$ for any finite $\tau_{0}$.

Finally, consider $\R[][(2)]$. With the change of variable $s \to r = s/\Delta\tau$, \eqref{RMW2 Detector at Rest} reads
\begin{align} \label{RMW2 Collapse Trajectory}
	\R[][(2)]
		&= \frac{2 \Delta\tau \sqrt{w'(\tau)}}{\pi^{2} \Delta{t}_{\mathrm{ret}}(\tau)}
			\int_{0}^{1} \dd{r} \cos(\omega r \Delta\tau)
				\frac{\sqrt{w'(\tau-r\Delta\tau)}}{\Delta{t}_{\mathrm{ret}}(\tau-r\Delta\tau)} 
\end{align}
Since $w'(\tau-r\Delta\tau) \leq 1$ and $\Delta{t}_{\mathrm{ret}}(\tau-r\Delta\tau) \geq 2d$, it follows that the integral factor in \eqref{RMW2 Collapse Trajectory} is bounded by a $\tau$-independent constant. At late times, the factor $\Delta\tau/\Delta{t}_{\mathrm{ret}}(\tau)$ converges to unity while $w'(\tau)$ vanishes exponentially. Hence, $\R[][(2)] = \smallO(1)$ as $\tau \to \infty$.

Combining the results, we see that only $\R[][(0)]$ has a non-vanishing contribution \eqref{RMW0 Collapse Trajectory Final} to the transition rate at late times when $\tau_{0}$ is strictly finite, regardless of which interaction Hamiltonian governs the detector-field interaction. The result \eqref{Late time transition rate for collapse model} hence follows.

\subsection{Asymptotically Early Switch-On Time ($H_{\mathrm{int}}^{\mathrm{no}}$ only)}

Now, consider a detector interacting via \eqref{Normally Ordered Interaction Hamiltonian} and is switched on in the asymptotic past. Using the fact that the last term in \eqref{RMW0 Detector at Rest} is $\BigO(1/\Delta\tau)$ as $\tau_{0} \to -\infty$, it follows that $\R[][(0)]$ is now given only by the first two terms of \eqref{RMW0 Collapse Trajectory 1}. Since the last term in \eqref{RMW0 Collapse Trajectory 1} vanishes as $\tau \to \infty$ when $\tau_{0}$ is finite, the late time limit of $\R[][(0)]$ is again given by \eqref{RMW0 Collapse Trajectory Final}.

For $\R[][(1)]$, we can always choose to split the integration interval of \eqref{RMW1 Detector at Rest}, which is now $[0,\infty)$, into $[0,\tau-\tau') \cup [\tau-\tau',\infty)$ where $\tau' < \tau$ is some arbitrary constant. Choosing $\tau' < -2d$, in particular, and start analysing at $\tau$ such that $\tau-\tau' > \Delta{t}_{\mathrm{ret}}(\tau)$, we may then use a similar method as above to show that the contribution from the interval $[0,\tau-\tau')$ vanishes at late times. The contribution from the other interval $[\tau-\tau',\infty)$ is given by
\begin{align}
	\Delta\R[][(1)]
		&= \frac{1}{\pi^{2}} \int_{\tau-\tau'}^{\infty} \dd{s}
			\frac{\cos(\omega s) \sqrt{w'(\tau) w'(\tau-s)}}{(s - \Delta{t}_{\mathrm{ret}}(\tau))(s + \Delta{t}_{\mathrm{ret}}(\tau-s))},
\end{align}
where we have set $\epsilon = 0$ since the integrand has no singularity. From $\Delta{t}_{\mathrm{ret}}(\tau-s) \geq 2d$, it follows that $s + \Delta{t}_{\mathrm{ret}}(\tau-s) \geq s + 2d$. Together with $\big| \cos(\omega s) \sqrt{w'(\tau) w'(\tau-s)} \big| \leq 1$, we have
\begin{align}
	\abs{\Delta\R[][(1)]}
		&\leq - \frac{1}{\pi^{2} (2d + \Delta{t}_{\mathrm{ret}}(\tau)) } \ln(\frac{1-\frac{\Delta{t}_{\mathrm{ret}}(\tau)}{\tau-\tau'}}{1+\frac{2d}{\tau-\tau'}}),
\end{align}
which implies that $\Delta\R[][(1)]$ vanishes as $\tau \to \infty$.

Combining the results, we see that \eqref{Late time transition rate for collapse model} remains valid in the case of asymptotically early switch-on time when the detector interacts via the normally-ordered interaction Hamiltonian $H_{\mathrm{int}}^{\mathrm{no}}$ given by \eqref{Normally Ordered Interaction Hamiltonian}.

\bibliographystyle{unsrt-v3.1-initial-firstname-hyperlinked-refs}
\bibliography{moving-mirror-thermal-radiation-references-v3.1}

\end{document}